\documentclass[twocolumn,letterpaper,aps,prc,superscriptaddress,nofootinbib,showpacs,floatfix]{revtex4-1}
\usepackage{graphicx}
\usepackage{comment}
\usepackage{setspace}
\usepackage{amsmath}
\usepackage{amssymb}
\usepackage[textwidth=16cm,textheight=22cm]{geometry}
\usepackage{color}
\usepackage{indentfirst}
\usepackage{xspace}
\usepackage{hyperref}
\usepackage{verbatim}
\usepackage{epstopdf}

\usepackage{lineno}

\begin{document}
\title{Effect of Rope Hadronisation on Strangeness Enhancement  in p$-$p collisions at LHC energies.}

\author{Ranjit Nayak}
\email{ranjit@phy.iitb.ac.in}
\affiliation{Indian Institute of Technology Bombay, Mumbai, 
  India-400076}
  
\author{Subhadip~Pal}
\email{sp15ms159@iiserkol.ac.in}
\affiliation{Indian Institute of Science Education and Research Kolkata , Mohanpur, 
  India-741246}

\author{ Sadhana~Dash }
\email{sadhana@phy.iitb.ac.in}
\affiliation{Indian Institute of Technology Bombay, Mumbai, 
  India-400076}

\begin{abstract}
The  p$-$p collisions at high multiplicity at LHC show small scale collective effects similar to that observed in heavy ion collisions such as 
enhanced production of strange and multi-strange hadrons, long range azimuthal correlations, etc. The observation of strangeness enhancement in p$-$p collisions at at $\sqrt{s}$ = 7 TeV  and 13 TeV as measured by ALICE experiment is explored using Pythia8 event generator within the framework of microscopic rope hadronization model which assumes the formation of ropes due to overlapping of  strings in high multiplicity environment.
The transverse momentum ($p_{T}$) spectra shape and its hardening with multiplicity is well described by the model. The mechanism of formation of ropes also described the observed experimental strangeness enhancement for higher multiplicity classes in p$-$p collisions at 7 TeV and 13 TeV.  The enhancement with multiplicity is further investigated by studying the mean $p_{T}$ ($<p_{T}>$) and the integrated yields ($<dN/dy>$ ) of strange and multi-strange hadrons and comparing the predictions to the measured data at  LHC for 7 TeV and 13 TeV. 
 \end{abstract}

\maketitle

\section{Introduction}

The recent observation of enhanced production of strange and multi-strange 
hadrons in p$-$p collisions at at $\sqrt{s}$ = 7 TeV with high final state multiplicity  as measured 
by ALICE experiment \cite{nature} and long range azimuthal correlations measured by CMS and ATLAS experiment \cite{cms1,cms2,atlas} 
have generated a lot of interest in small systems. These observations mimic features present in 
deconfined matter formed in heavy ion collisions and are manifestations of Quark gluon Plasma (QGP) dynamics.
The QGP observables are studied as a function of charged particle density ($dN_{ch}/d\eta$) in smaller systems
and the size of the effects seem to agree with heavy ion collisions for similar charged particle density.
Therefore, the possibility of formation of a mini-QGP in small systems for high multiplicity is also being predicted.  

An enhanced production of strange and mutli-strange hadrons in heavy ion collisions 
has long been predicted to be a signature of the formation of the QGP medium \cite{rafelski}. The strange valence quark is not present in the initial state of colliding nuclei and hence are produced by hard partonic 
scattering processes such as flavor excitation and flavor creation in
initial stages of collision and therefore the strange hadrons are
predominantly produced in high $p_{T}$ region. However, in the low
$p_{T}$ region, where the perturbative description fails, the
production of strange hadrons is suppressed compared to the light
quark ($u$ and $d$) hadrons owing to the relatively heavier mass of
strange quarks. In heavy ion collisions, the enhanced production of
strange particles in central and mid-central collisions have been
attributed to the abundance of strange and anti-strange quarks in the deconfined QGP medium \cite{koch,rafelski}. The heavy ion data was also successfully described using statistical thermal models assuming a 
grand canonical ensemble approach.In peripheral collisions, the
strangeness production is similar to that what observed in p$-$p
collisions and was attributed to strangeness canonical suppression. 
The measurement by ALICE experiment might indicate towards the formation of mini-QGP like
system in p$-$p collisions \cite{nature}. Many alternative explanations based on increased and complex interactions 
among partons in the fragmentation phase of the Quantum Chromodynamics (QCD) based string hadronization models were put forward to explain some  of the  experimental  observations. The description of these dynamic interactions among partons do not assume a formation of 
deconfined and thermalized plasma state. One of the mechanism in QCD based String fragmentation is rope hadronization which is quite
successful in describing the enhancement \cite{bielrich1,dipsy}.  
In this work, we study the strangeness enhancement observed in p$-$p
collisions at $\sqrt{s}$ = 7 TeV and 13 TeV using rope hadronization mechanism of 
PYTHIA \cite{bielrich1,bielrich2} and compare with the ALICE measurement\cite{nature}. 
The predictions of some observables like mean transverse momentum ($\langle p_{T} \rangle$) and mean integrated yields($ \langle dN/dy \rangle$) of strange hadrons have also been studied to understand the dependence of strange particle production on beam energy.
\section{Rope Hadronization}
In high energy p$-$p collisions, the particle production can be
simplified into two broad steps namely initial hard scattering leading
the production of partons and the subsequent hadronization of the
initial parton configuration \cite{lund1,lund2}. The generation of
partons and the partonic level activity involving multiple
parton-parton interactions, the initial and final state radiations, and 
the activity of beam remnants can be approximately described by
perturbative QCD while the fragmentation of the final parton
configuration to observable hadrons is completely
non-perturbative in nature. The understanding of the later part depends on statistical
parametrization of experimental data, realistic modeling,  parameter tuning etc.
One of the fragmentation models namely The Lund string model has been 
extremely successful in describing the hadronization process which
envisages  stretched color flux tubes between two partons leading to
linear confinement via massless relativistic string.  As the potential
energy in the string increases, the quark-antiquark pair move apart
and the string breaks producing a new quark-antiquark pair. The hadrons are 
formed by combining these quarks.
In high energy collisions, one can have many such overlapping strings
in a small transverse area due to multi-parton interactions  leading to
description of underlying event activity within the framework of rope
hadronization \cite{rope1,rope2}. Within this framework, the
overlapping colored  strings act coherently to form a  color rope which subsequently 
hadronize  with a higher effective string tension. The  breakup of strings with 
higher string tension produce more strange quarks and di-quarks resulting in enhanced 
production of baryons and strange hadrons.

\section{ PYTHIA Event generator  and Analysis }
PYTHIA is a standalone Monte-Carlo event generator for high
energy e$-$e, p$-$p and $\mu-\mu$ collisions and has been extensively
used to study LHC physics and compare its data. The routines are completely written 
in C++ (compared to its predecessor PYTHIA6 which was in Fortran) with some 
introduction of new physics and improvisations in the existing
processes at both partonic and hadronic level. The details of the physics processes and 
its implementation can be found in reference \cite{pythia8}. 
In this present work, the Monash 2013 tune of Pythia8.2 \cite{monash}
based on larger and recent set of LHC data has been used to
generate events for p$-$p collisions. The multi-parton interactions are
enabled with QCD-based  color reconnections where one obtains different quark junctions 
due to reconnections of hadronizing strings \cite{rope3,rope2}. The analysis was also checked with MPI based CR scheme
where one obtains smaller string length due to fusion of color flow between partons belonging to
different MPI system. The subsequent hadronization  is studied within the 
framework of formation of color ropes. The results (only shown for QCD-based CR scheme) are 
also compared to the scenario where the mechanism of rope hadronization is not included.

The analysis is done by generating 50 Million inelastic non-diffractive events with soft
QCD processes for p$-$p collisions at $\sqrt{s}$ = 7 TeV and 13 TeV with rope hadronization and without it. 
As the aim of this work is to see whether the strangeness enhancement as observed by ALICE experiment can be explained with 
the mechanism of rope hadronization, the multiplicity classes were determined by dividing the event sample into ten different event classes 
based on the total charged multiplicity within the acceptance of ALICE V0 detectors. The yield of strange ($K^{0}_{S}$ and $\Lambda + \overline{\Lambda}$) and multi-strange hadrons ($\Xi^{-} + {\overline{\Xi}}^{+}$ and $\Omega^{-} + {\overline{\Omega}}^{+}$) was obtained for $|y| < 0.5$ for different 
multiplicity classes. The mean pseudorapidity density of charged particles, $\langle dN_{ch}/d\eta \rangle$ was also
estimated for $|\eta| < 0.5$. The multiplicity class and the corresponding $<dN_{ch}/d\eta>$ range as used by ALICE experiment for the strangeness enhancement study is shown in Table \ref{tab1}. 
\begin{table}
\centering
\begin{tabular}{c c c c }
\hline
Multiplicity  &  $\langle dN_{ch}/d\eta \rangle $ & Multiplicity   & $\langle dN_{ch}/d\eta \rangle$ \\
class & &class&\\
\hline 
\hline
I &  21.5 $\pm$ 0.6   & VI  &  8.45 $\pm$ 0.25 \\
II  &  16.5 $\pm$ 0.5 & VII & 6.72 $\pm$ 0.21\\
III &  13.5 $\pm$ 0.4 & VIII  & 5.40 $\pm$ 0.17\\
 IV  & 11.5 $\pm$ 0.3  & IX & 3.90 $\pm$ 0.14\\
 V   &  10.1 $\pm$ 0.3  & X & 2.26 $\pm$ 0.12\\
\hline 
\end{tabular}
\caption{ \label{tab1} The values of  ${\langle dN_{ch}/d\eta \rangle }_{\eta < 0.5}$  for different multiplicity classes as used
by ALICE experiment for p$-$p collisions at $\sqrt{s}$ = 7 TeV \cite{nature}.}
\end{table}
\begin{figure*}

\includegraphics[scale=0.7]{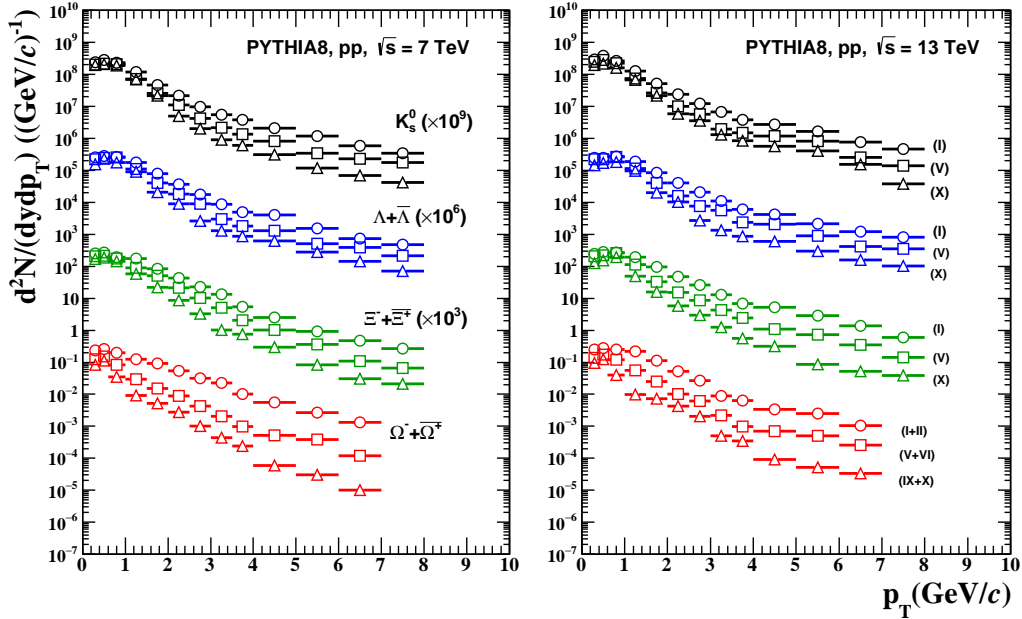}
\caption{(Color online) $p_{T}$ distribution of strange and mutli-strange hadrons in p$-$p collisions at 7 TeV (left panel) 
and 13 TeV (right panel) for $|y| < 0.5$.}
\label{fig1}
\end{figure*}
\begin{figure*}
\begin{center}
\includegraphics[scale=0.8]{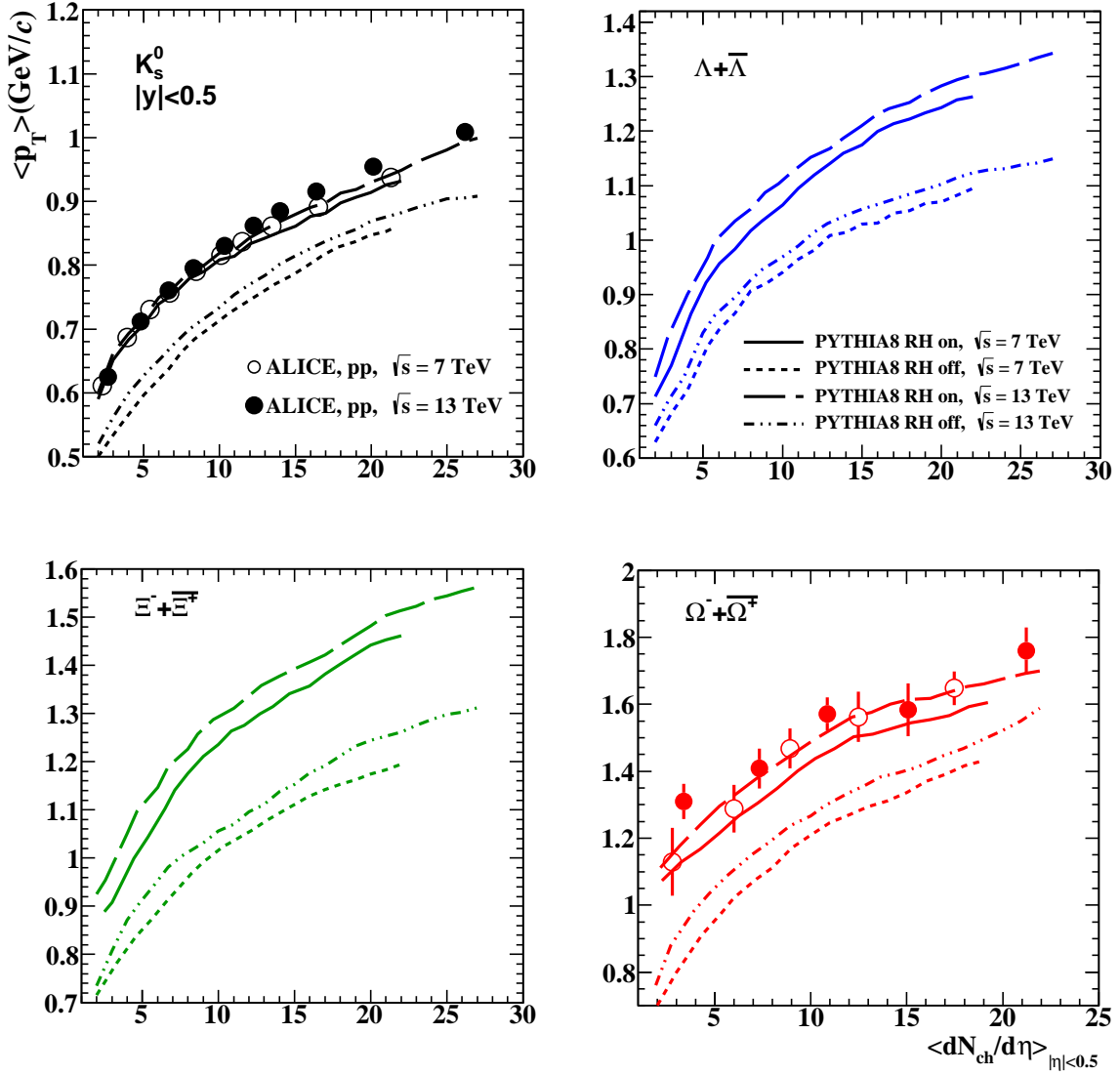}
\caption{(Color online) $\langle p_{T} \rangle$ as a function of  multiplicity ( $<dN_{ch}/d\eta>$) for strange hadrons predicted by 
PYTHIA8  with and without rope hadronisation in p$-$p collisions at 7 TeV and 13 TeV. The PYTHIA estimations at 7 TeV  and 13 TeV 
are also compared to the available data as measured by ALICE experiment\cite{aliceproc} .}

\label{fig2}
\end{center}
\end{figure*}

\section{ RESULTS  AND COMPARISON TO DATA}

The analysis was carried out with the generated events using PYTHIA generator and the estimates of various observables 
were studied and compared to the existing data to understand the  mechanism of strange particle production in high multiplicity 
p$-$p collisions. The  $p_{T}$ spectra of $K^{0}_{S}$, $\Lambda$, $\Xi^{-}$ and $\Omega^{-}$)  for three different multiplicity classes are 
shown for p$-$p collisions at $\sqrt{s} = $ 7 TeV (left panel) and 13 TeV (right panel) with rope hadronization mechanism in Figure 1\ref{fig1}.  
One can observe that  the  $p_{T}$  spectra becomes harder with an increase in multiplicity as observed in data measured by ALICE experiment.
It is also important to note that the hardening is  conspicuous for higher mass particles. This feature is consistent with the onset of collectivity in Pb$-$Pb collisions. However, it is important to note that the hardening of the  spectra as well the hardening becoming more for higher
 mass particles is well reproduced by rope hadronization  mechanism where the formation of plasma or collectivity is not assumed.
 The rate of hardening for various multiplicity classes can be compared by  estimating the mean transverse momentum ($<p_{T}>$)
 as a function of multiplicity. Figure \ref{fig2} compares the  $<p_{T}>$ of $K^{0}_{S}$ and $\Omega^{-}$ as measured by ALICE experiment with
  the predictions of Pythia8 with (and without ) rope hadronization. The estimations for $\Lambda$ and $\Xi$ is also shown.
  It is worth noting that the evolution of mean $<p_{T}>$ of strange hadrons with event  multiplicity  is well reproduced by rope hadronization 
  for both 7 TeV and 13 TeV. The PYTHIA tune where there is no formation of color ropes do not describe the evolution  satisfactorily and the discrepancy becomes more for strange hadrons with higher mass. 
 
\begin{figure}
\includegraphics[scale=0.38]{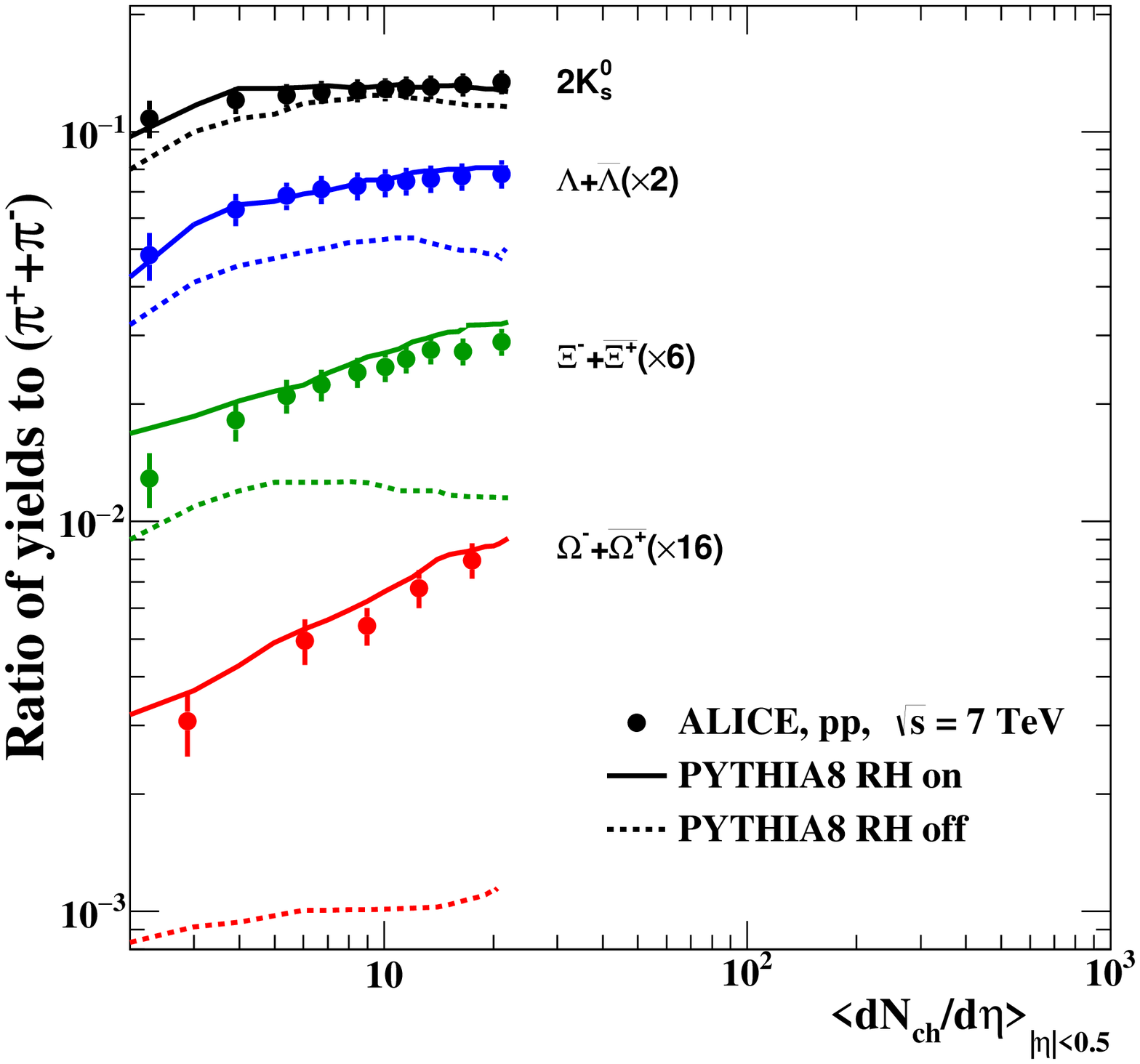}
\includegraphics[scale=0.38]{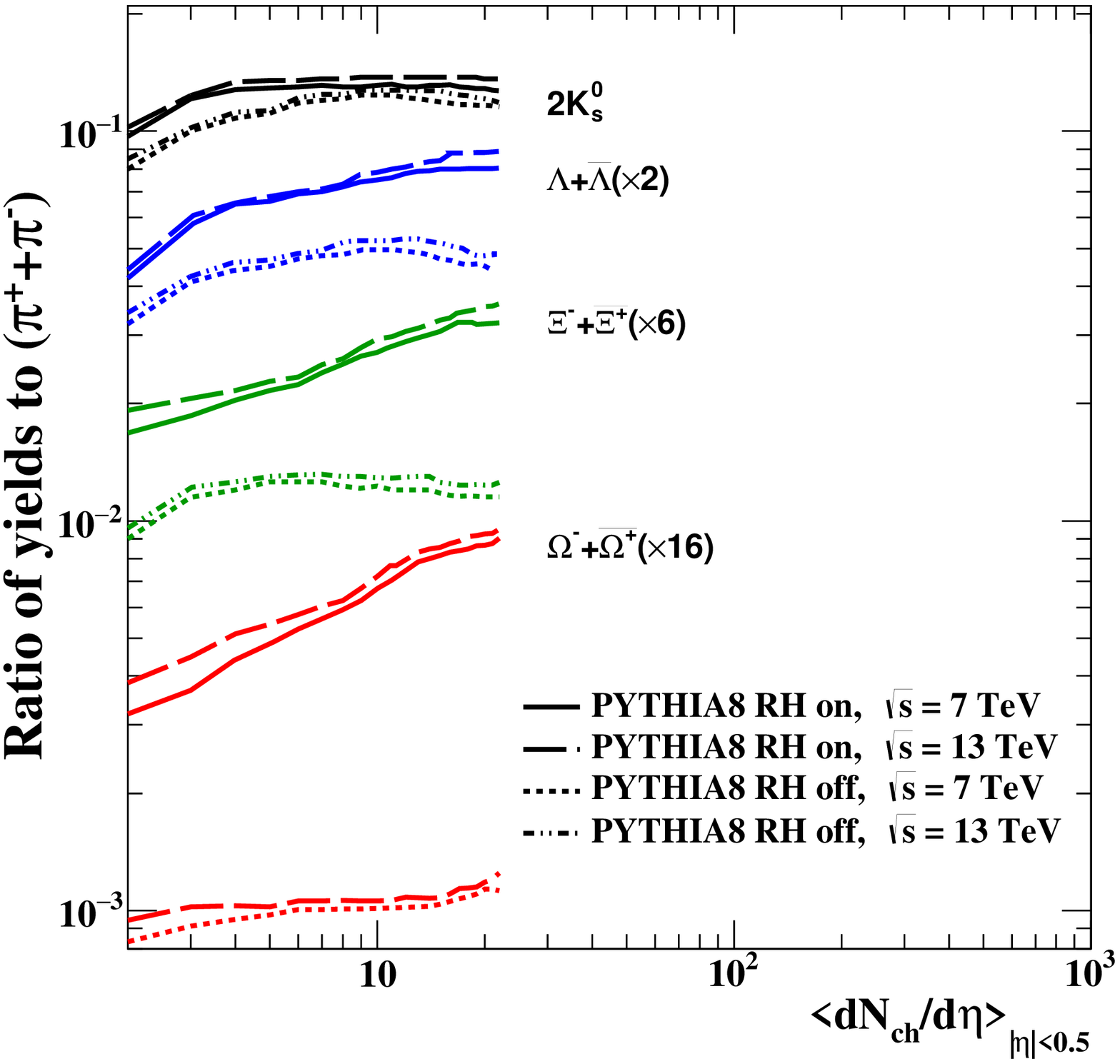}
\caption{(Color online) ({\bf Upper Panel}) Ratio of yield of strange hadrons to pions as
  a function of $dN_{ch}/d\eta$ for $|\eta| < 0.5$. The solid
  markers are data as measured by ALICE experiment at LHC
  \cite{nature}. The solid and dotted lines are estimates of Pythia  with rope
  hadronization on and off, respectively. The data points and the Pythia 
  predictions are scaled for visibility.
  ({\bf Lower Panel}) The Pythia estimates of the ratio for p$-$p collisions at 13 TeV compared with 
  7 TeV.}
 
  
\label{fig3}
\end{figure} 
\begin{figure}
\includegraphics[scale=0.39]{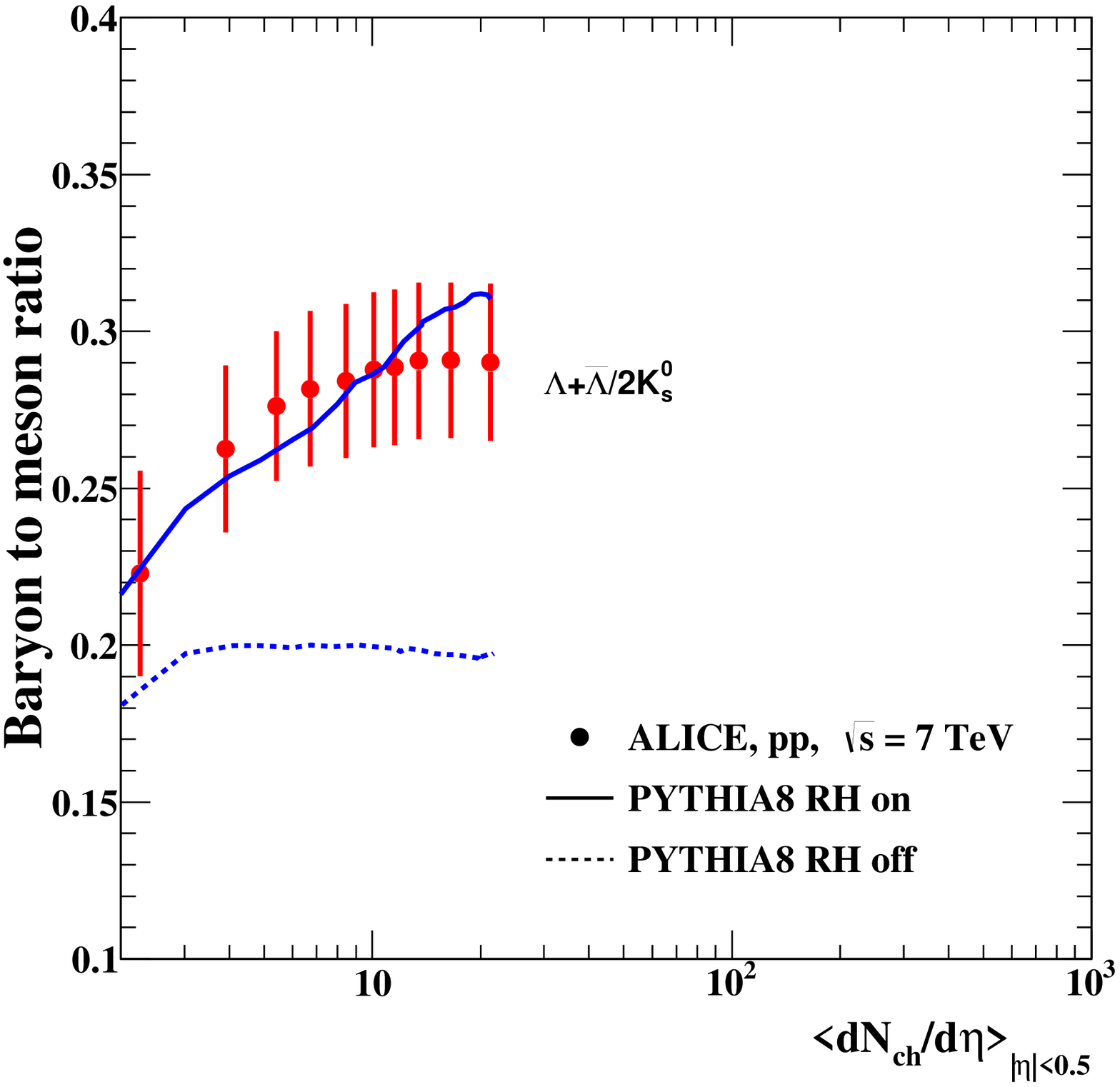}
\caption{(Color online) $\Lambda/K^{0}_{s}$ ratio as a function of $<dN_{ch}/d\eta>$}
\label{fig4}
\end{figure}

\begin{figure*}
\begin{center}
\includegraphics[scale=0.7]{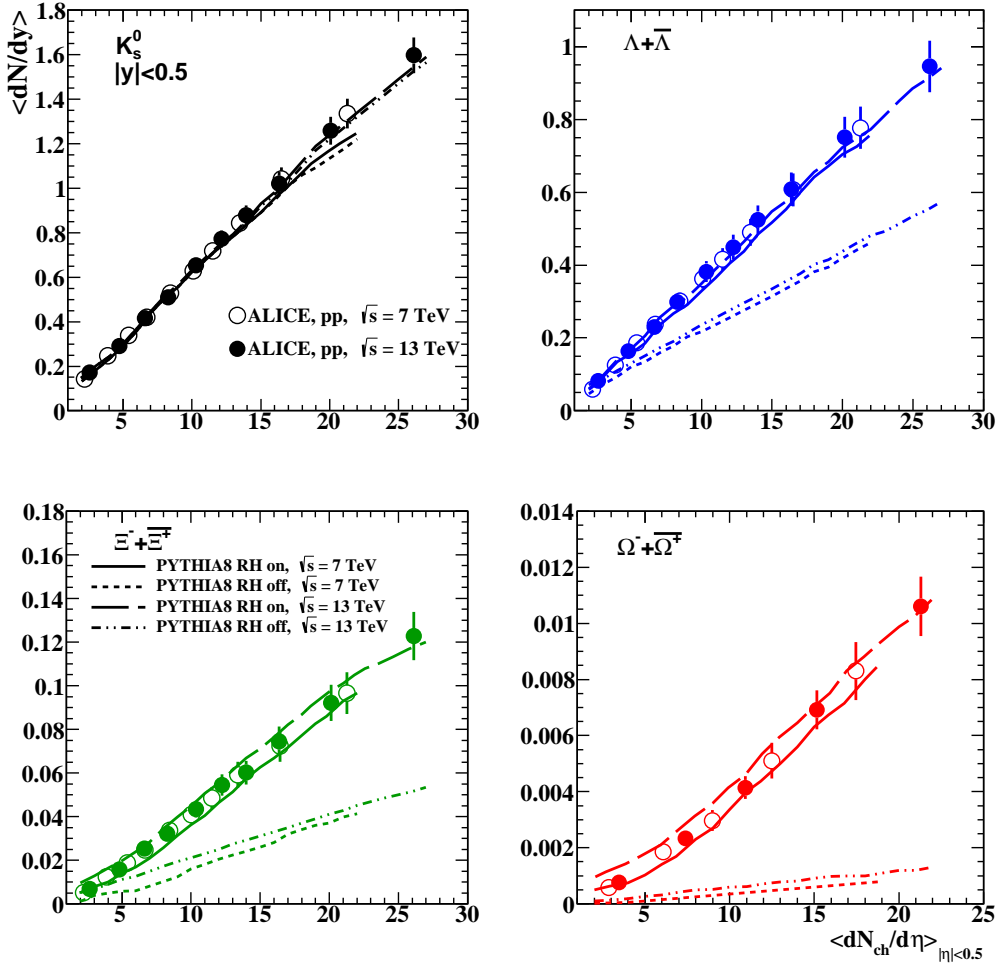}
\caption{(Color online) The PYTHIA8 predictions of  $p_{T}$ integrated yields for $K^{0}_{S}$, $\Lambda + \overline{\Lambda}$ , $\Xi^{-} + {\overline{\Xi}}^{+}$ and $\Omega^{-} + {\overline{\Omega}}^{+}$ as a function of  $<dN_{ch}/d\eta>_{|\eta| < 0.5}$  using rope hadronisation. The estimations are also compared to the data measured by ALICE experiment at LHC \cite{aliceproc}.}

\label{fig5}
\end{center}
\end{figure*}

The enhanced production of strange hadrons is quantified by measuring the ratio of strange hadron yield with respect
to the yield of pion (non-strange particle) in the same acceptance. Figure~\ref{fig3} compares the enhancement of strange and
multi-strange hadrons as a function of multiplicity as measured by ALICE experiment to the values obtained with PYTHIA generator.
As can be seen in the figure, the mechanism of rope hadronization (with color reconnections) in PYTHIA  describes the observed 
experimental strangeness enhancement in p$-$p collisions at 7 TeV while the estimations without rope hadronization do not show the same.
A similar result where the charged particle multiplicity was measured in the forward region with rope hadronization mechanism can be found here\cite{}.  The predicted enhancement for p$-$p collisions at 13 TeV and its comparison with 7 TeV is shown in the lower 
panel of figure \ref{fig3}. There is no significant difference in enhancement  obtained for 13 TeV for similar multiplicity classes.
In a previous study,  it was shown that the yield ratios $\Lambda/K^{0}_{S}$  did not vary much as a function of multiplicity for 
p$-$p collisions at 7 TeV \cite{nature} and the observation was not explained by PYTHIA 8 predictions. However, one observes that the variation of ratio, $\Lambda/K^{0}_{S}$ with multiplicity is described well when one incorporates the mechanism of  rope formation as shown in Fig \ref{fig4}. The observation also indicated that the observed strangeness enhancement is not due to the hadronic mass or specie of the particle but can be attributed to enhanced production of strange particles  due to  possible formation of color ropes.
The enhancement was further investigated by comparing the multiplicity dependence of 
the integrated yields of strange hadrons ($K^{0}_{S}$, $\Lambda$, $\Xi^{-}$ and $\Omega^{-}$) in p$-$p collisions at  7 TeV and 13 TeV . 
This PYTHIA8 predictions are compared with the available data in Fig. \ref{fig5}. One observes that the mechanism of rope hadronization essentially captures the evolution of enhancement with multiplicity and the estimates are in good agreement with the data.  
One can also conclude that the enhancement is  driven by the event activity rather than the variation in centre of mass energies. 
It can be seen that the evolution of integrated yield with $<dN_{ch}/d\eta>$ for $K^{0}_{S}$ can be described by both PYTHIA tunes without 
the formation of ropes while they under predict the data for strange baryons. However, there is a good agreement with the data (within the experimental errors ) when one includes the rope hadronization scenario.

\section{Summary}
The recent observation of enhanced production of strange and multi-strange hadrons in p$-$p collisions  at  $\sqrt{s} =$ 7 TeV and 13 TeV as 
measured by ALICE experiment is well described by the microscopic model of rope hadronization implemented in Pythia8.  The 
$p_{T}$ spectra  of the strange hadrons ($K^{0}_{S}$, $\Lambda + \bar{\Lambda}$ , $\Xi^{-} + \bar{\Xi^{+}}$ and $\Omega^{-} + \Omega^{+}$) for different multiplicity classes were observed to harden with an increase in multiplicity and the hardening was conspicuous 
for particles with higher mass. This was further investigated by studying the  multiplicity dependence of $\langle p_{T} \rangle$ and $\langle dN/dy \rangle$ of the strange hadrons. The evolution of these observables was nicely described by the mechanism of rope hadronization while the 
one without this mechanism did not describe the data. The  rope hadronization  which do not assume the formation of any thermalized plasma 
state also described the strangeness enhancement (which is quantified by the ratio of strange hadrons to pions ) observed in data for p$-$p collisions at $\sqrt{s}$ = 7 TeV and 13 TeV. The observed strangeness enhancement  seems to saturate at higher multiplicities for both the collision energy.  It will be interesting to observe the application of rope hadronization mechanism to other particle production scenario
involving higher mass particles such as heavy flavor hadrons and resonance particles. 

\section{Acknowledgements}
The authors would like to thank Christian Bierlich for his valuable 
suggestions on rope hadronisation. Sadhana Dash would like to thank the 
Department of Science and Technology (DST), India for supporting the present work.

\end{document}